\documentclass{jpsj3}
\usepackage{txfonts}
\usepackage{color}
\usepackage{bm}

\title
{Electromagnetic Responses of Vortex Lattices in Unconventional Superconductors
}

\author{Ryusuke Ikeda and Yuto Yokota}

\inst{Department of Physics, Graduate School of Science, Kyoto University, Kyoto 606-8502, Japan
} %\\

\abst{ 
The electro-magnetic responses of ordered vortex lattices in unconventional superconductors are studied in a high field approximation. In the cases with a vortex lattice formed within the lowest Landau level of the superconducting order parameter (OP) such as a conventional $s$-wave paired system with a single OP and a nonchiral spin triplet paired one with multiple components of OPs, the vanishing of the superfluid stiffness for a gauge field disturbance perpendicular to the applied uniform magnetic field is found to be ensured only for the vortex lattice structures minimizing the free energy. The notion of the vanishing superfluid stiffness ensured by minimization of the free energy is found to be satisfied in a more complex $d$-wave pairing case where the vortex lattice in lower fields has an anisotropic structure deviated from the six-fold hexagonal symmetry. Interestingly, such an anisotropy in the vortex lattice structure of a $d$-wave paired superconductor is reflected not in the resulting vortex flow conductivities obtained after minimizing the free energy but in the elastic energy describing the harmonic fluctuation around the vortex lattice state. Relevance of the obtained results to the vortex pinning effects are discussed. 
}

\begin{document}

%\keyword{}

\maketitle

\section{Introduction}

It is well accepted that, in a perfectly clean superconductor, a single vortex excitation flows under a homogeneous current \cite{dG}. In addition, it is believed \cite{CM,Kopnin,TroyDorsey,com} that a vortex lattice in a pinning-free system also flows under a homogeneous current, like the vortex flow of a single vortex mentioned above, irrespective of whether the lattice structure is perfectly ordered or not. However, it is not necessarily clear whether this conventional wisdom is valid : Intuitively, a single vortex flow mentioned above can be regarded as an extremely simplified picture on the resistive behavior in the uncorrelated vortex {\it liquid}, while the correlated vortex lattice or solid is the ordered phase which occurs only through the freezing phase transition of the disordered vortex liquid. It has been found in our recent study \cite{Yokota} that the superfluid stiffness for a current perpendicular to the magnetic field is nonzero in vortex lattices specifically occurring in superconductors with strong paramagnetic pair-breaking. Then, it should be questioned under what condition the rigid flow motion of an ordered vortex solid under a uniform current is satisfied. 

In the present work, the superfluid stiffness (or, helicity modulus) \cite{RI95,Feigel} and the electric conductivities in the mean field vortex lattice phase are examined for several typical superconductors in the high field approximation. First, we examine the conventional vortex lattice in $s$-wave paired superconductors and a simple model of a spin-triplet superconductor described by multiple components of the complex scalar order parameters (OPs). These two models are common in that the vortex lattices can be described in terms only of the lowest Landau level (LL) of the OPs. It is found that their vortex flow responses are ensured only for the vortex structure minimizing the free energy. Our analysis is extended to the vortex lattice in a $d$-wave superconductor which has some deviation in structure from the hexagonal six-fold symmetry. It is found even in this case including effects of higher LLs that minimizing the free energy of the lattice structure is needed to realize the vortex flow response and to keep the vortex flow conductivities in the plane perpendicular to the magnetic field isotropic. Through these results, we argue that the flow response of the vortex lattice essentially differs from the corresponding response of a single vortex. 

The present paper is organized as follows. In sec.II, responses of the conventional vortex lattice are reviewed, and the corresponding issue of a simple model consisting of multiple components of the OPs are examined in sec.III. The case of the $d$-wave superconductor is examined in sec.IV, and a summary and comments on vortex pinning effects are given in sec.V. Mathematical details necessary for our analysis in the main text are explained in Appendix.

\section{Superfluid Stiffness in Vortex Lattice of Conventional Superconductor}

We start from the conventional Ginzburg-Landau (GL) hamiltonian expressed by a single complex scalar OP $\Delta$ 
\begin{equation}
{\tilde {\cal H}}_1 = \frac{{\cal H}_1}{N(0)} = \int d^3\bm{r} \biggl[ - \varepsilon_{0} |\Delta|^2 + \xi_0^2 \,|\bm{\Pi} \Delta|^2 + \frac{g}{2} |\Delta|^4 \biggr], 
\label{GLsinglet}
\end{equation}
where $N(0)$ is the density of states on the Fermi surface of the conduction electrons, $\Delta$ is the superconducting OP, 
$\xi_0$ is the GL coherence length, 
\begin{equation}
\bm{\Pi} = - i \bm{\nabla} + \frac{2 \pi}{\phi_0} \bm{A}, \qquad  \bm{A} = \bm{A}_{\rm {ex}} + \delta \bm{A} 
\label{Piope}
\end{equation}
is the gauge-invariant gradient operator, $\phi_0 = \pi c \hbar/|e|$ is the flux quantum for the charge $2e$, and $\varepsilon_0 = {\rm ln}(T_c(0)/T)$ with the zero field superconducting transition temperature $T_c(0)$ which is always positive in any situation of our interest below. In eq.(\ref{Piope}), 
the gauge field was divided into the external one $\bm{A}_{\rm {ex}}(\bm{r}) = B y {\hat e}_x$ satisfying $\bm{\nabla} \times \bm{A}_{\rm {ex}} = - B {\hat e}_z$ and the disturbance $\delta \bm{A}$. The component of the gauge field spatially varying on the length scale of the period of the vortex lattice will be neglected by focusing on the type II limit. 

The raising and lowering operators of LLs of the superconducting OP are given by 
\begin{equation}
{\hat a}^\dagger = \frac{1}{\sqrt{2}} \left( - i \frac{\partial}{\partial {\overline x}} + {\overline y} - \frac{\partial}{\partial {\overline y}} \right) 
\end{equation}
and 
\begin{equation}
{\hat a} = \frac{1}{\sqrt{2}} \left( - i \frac{\partial}{\partial {\overline x}} + {\overline y} + \frac{\partial}{\partial {\overline y}} \right)
\end{equation}
, respectively, and they satisfy the commutation relation ${\hat a} {\hat a}^\dagger - {\hat a}^\dagger {\hat a} = 1$, where ${\overline {\bm{r}}} = \bm{r}/r_B$, and $r_B = \sqrt{\phi_0/(2 \pi B)}$. 

The LL eigen function is constructed in several ways based on the lowest ($n=0$) Landau level (LL) eigen function $\varphi_0({\bm{r}}|0)$ : 
\begin{eqnarray}
\varphi_n({\bm{r}}|0) &=& \frac{({\hat a}^\dagger)^n}{\sqrt{n!}} \varphi_0({\bm{r}}|0) \nonumber \\ 
&=& \frac{1}{\sqrt{n!}} \frac{\partial^n}{\partial t^n} e^{t^2/2} \varphi_0({\bm{r}} - \sqrt{2} r_B t {\hat y}|0) |_{t \to 0},  
\label{nthLL}
\end{eqnarray}
where the second representation follows from the use of the generating function of the Hermite polynomial \cite{Eilenberger}. The complete set of LLs is constructed in the manner \cite{Eilenberger}
\begin{equation}
\varphi_n(\bm{r}|\bm{r}_0) = e^{i {\overline y}_0 {\overline x}} \varphi_n(\bm{r}+\bm{r}_0|0)
\label{Gert}
\end{equation}
in terms of the continuous vector $\bm{r}_0$. 
The lowest LL eigen function describing a general periodic vortex lattice structure, $\varphi_0(\bm{r}|0)$, takes the form 
\begin{equation}
\varphi_0(\bm{r}|0) = \left(\frac{k^2}{\pi} \right)^{1/4} \sum_n \, e^{i \pi R n^2} \, e^{i k n {\overline x} - \frac{1}{2} ({\overline y} + k n)^2}. 
\label{Abrikosovsol}
\end{equation}
which satisfies $\langle [\varphi_n(\bm{r}|0)]^*  \varphi_m(\bm{r}|0) \rangle_s = \delta_{m n}$, where $\langle \,\,\, \rangle_s$ denotes the space average.  

Near $H_{c2}(T)$-line, the conventional vortex lattice in the $s$-wave case is well described by focusing on the lowest LL and based on (\ref{Abrikosovsol}). With decreasing $B$ and leaving from $H_{c2}(T)$-line, higher LLs with indices of multiples of six begin to contribute to describing the superconducting OP, and the expansion parameter for controlling the weight of those higher LLs is \cite{Lasher}
\begin{equation}
\frac{|\varepsilon|}{h} = \frac{H_{c2}(T) - B}{B}, 
\label{Lasherexp}
\end{equation}
where 
$\varepsilon = \varepsilon_0 - h$, and $h=B/H_{c2}(0)$. 
Then, by setting the mean field solution in high field approximation in the 
form 
\begin{equation}
\Delta_{0} = A_0 \varphi_0(\bm{r}|0), 
\label{Abrikosov0}
\end{equation}
the mean field equation resulting from minimizing with respect to $|A_0|^2$ is 
\begin{equation}
- \varepsilon + g |A_0|^2 \langle 0, 0|0, 0 \rangle = 0 .
\label{MFEQ0}
\end{equation} 

%%%%%%%%%%%%%%%%%%% 
\begin{figure}
\scalebox{0.3}[0.3]{\includegraphics{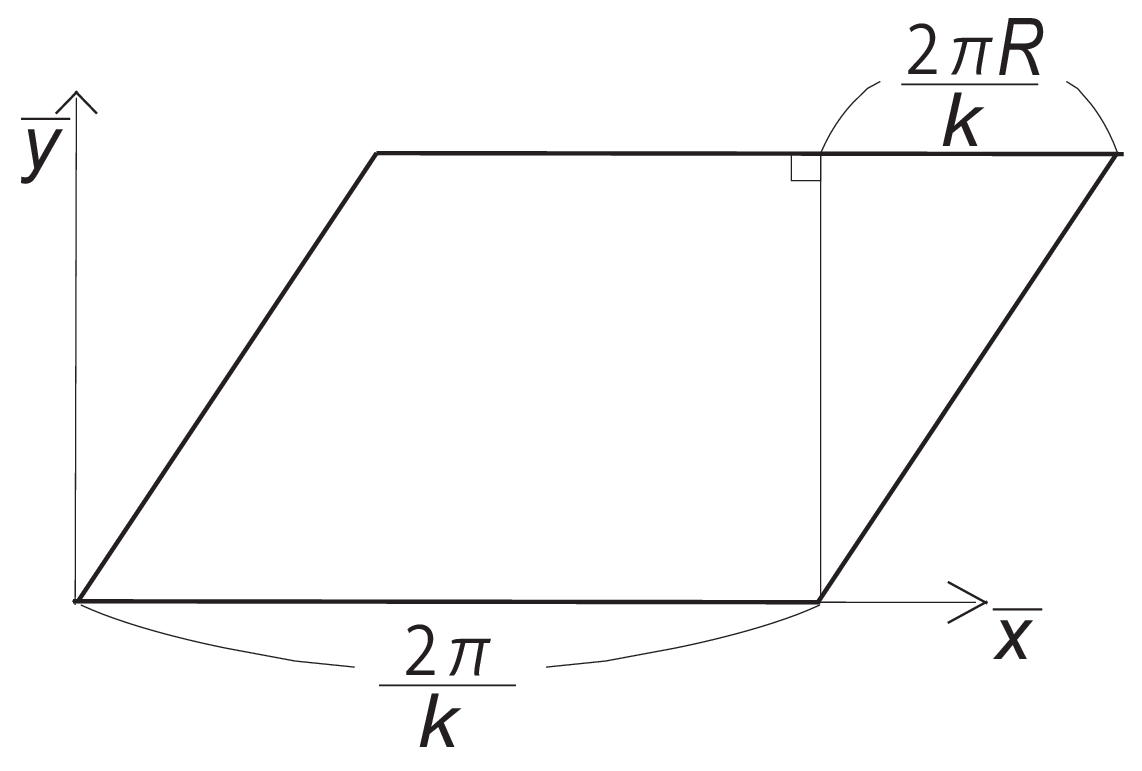}}
\caption{Parallelogram expressing the unit cell of the conventional vortex lattice. }
\label{fig.1}
\end{figure}
%%%%%%%%%%%%%%%%%

Next, the so-called Abrikosov factor \cite{AAA}
\begin{equation}
\beta^{(1)}_{\rm A} = \langle |\varphi_0(\bm{r}|0)|^4 \rangle_s \equiv \langle 0,0|0,0 \rangle , 
\label{Abrikosovfactor1}
\end{equation}
describing the lattice structure, is determined by minimizing $\beta^{(1)}_{\rm A}$ with respect to $k$ in the manner 
\begin{equation}
\partial \langle 0,0|0,0 \rangle \equiv k^2 \frac{\partial \langle 0,0|0,0 \rangle}{\partial k^2} = 0. 
\label{LLLmini}
\end{equation}
Here, we have focused on the structures with reflection symmetry in which $R=1/2$ or $R=0$ (see Fig.1). 

Let us turn to examining the superfluid stiffness defined by \cite{RI95} 
\begin{equation}
\Upsilon_{ij} = \frac{\delta^2 F(\delta \bm{A})}{\delta A_i \delta A_j} \biggr|_{\delta \bm{A}=0}, 
\label{ros}
\end{equation}
where $F(\delta \bm{A})$ is the free energy functional under the gauge disturbance $\delta \bm{A}$. To examine this quantity in the mean field ordered state, the only OP fluctuations we need to incorporate to obtain (\ref{ros}) are those coupling to $\delta \bm{A}$ in the GL Hamiltonian. However, we only have to include them at the harmonic level, since the response quantities in the mean field approximation are not accompanied by the thermal energy $k_{\rm B} T$, where $k_{\rm B}$ is the Boltzmann constant. 
The harmonic fluctuation contribution to the GL Hamiltonian (\ref{GLsinglet}) takes the form ${\cal H}_{\Delta} + {\cal H}_{A}$, 
where 
\begin{eqnarray}
{\tilde {\cal H}}_{\Delta} &=& \int d^3\bm{r} \left[ - \varepsilon_0 |\delta \Delta|^2 + h \delta \Delta^* (2{\hat a}^\dagger {\hat a} + 1) \delta \Delta + \frac{g}{2} \left( 4 |\Delta_{\rm MF}|^2 |\delta \Delta|^2 + ( \, (\Delta_{\rm MF}^*)^2 (\delta \Delta)^2 + {\rm c.c.} \, ) \right) \right], \nonumber \\
{\tilde {\cal H}}_{A} &=& \int d^3\bm{r} \biggl[ \delta {\tilde A}_+ \delta {\tilde A}_- |\Delta_{\rm MF}|^2  + \sqrt{\frac{h}{2}} ( \, \delta {\tilde A}_+ [({\hat a} \delta \Delta)^* \Delta_{\rm MF} + ({\hat a} \Delta_{\rm MF})^* \delta \Delta + [\Delta_{\rm MF}]^* {\hat a}^\dagger \delta \Delta \nonumber \\ 
&+& [\delta \Delta]^* {\hat a}^\dagger \Delta_{\rm MF}] + {\rm {c.c.}} ) 
\biggr], 
\label{harmgl1}
\end{eqnarray}
where $\delta {\tilde A}_\pm = \delta {\tilde A}_x \pm i \delta {\tilde A}_y$,  and  
\begin{equation}
\delta {\tilde {\bm{A}}} = \frac{2 \pi}{\phi_0} \xi_0 \, \delta \bm{A}. 
\label{gaugefluct}
\end{equation} 
In the case of a vortex lattice formed in the $n=0$ LL, only the $n=1$ LL fluctuation of OP 
\begin{equation}
\delta \Delta_1 = A_0 \, a_1 \varphi_1(\bm{r}|0). 
\label{fluc1}
\end{equation}
couples to $\delta \bm{A}$ \cite{RI95}. 
By identifying (\ref{Abrikosov0}) and (\ref{fluc1}) with $\Delta$ and $\delta \Delta$ in eq.(\ref{harmgl1}), respectively, and using eq.(\ref{bra1}) in Appendix, the expression of $\delta {\cal H}$ is rewritten in the following form 
\begin{equation}
\delta {\tilde {\cal H}} = 2 h |A_0|^2 |{\tilde a}_1|^2 + \frac{|A_0|^4}{2} g \biggl[\langle 0,0|1,1 \rangle a_1^2 + {\rm {c.c.}} \biggr], 
\label{flowLLL}
\end{equation}
where 
\begin{equation}
{\tilde a}_1 = a_1 - \frac{1}{\sqrt{2h}} \delta {\tilde A}_+. 
\label{joseph01}
\end{equation}
The Josephson relation $\delta \bm{A} + \bm{B} \times \bm{s}=0$ \cite{Joseph} ensuring the vanishing of the superfluid stiffness corresponds \cite{RIjoseph} to ${\tilde a}_1=0$, where $\bm{s}$ is the uniform displacement vector of the vortex lattice. 
If the bracket $\langle 0,0|1,1 \rangle$ is zero, $F(\delta\bm{A})$ vanishes after integrating over $a_1$, implying the vanishing of ${\tilde a}_1$ and hence, of the superfluid stiffness. 
In fact, as is explained in Appendix, it is clearly seen in eqs.(\ref{derbra0}) and (\ref{derbrackets}) that, by using the Poisson summation formula, the bracket $\langle 0,0|1,1 \rangle$ satisfies the relation 
\begin{equation}
\langle 0,0|1,1 \rangle = \partial \langle 0,0|0,0 \rangle
\label{00partial}
\end{equation}
for any lattice structure with $R=0$ or $1/2$. Then, according to eq.(\ref{LLLmini}), the last term of (\ref{flowLLL}) vanishes, and it is concluded that $\Upsilon_{ij}$ is zero. In this way, the vanishing $\Upsilon_s$ consistent with the vortex flow response is obtained {\it only} for the lattice structure minimizing the mean field free energy. This fact suggests that the vortex flow response of a vortex lattice is {\it not} a trivial extension of the single vortex dynamics.

\section{Structural Transitions and Response in Vortex Lattices of a Two Component Superconductor}

To verify whether the finding in the preceding section that the vortex flow response of the conventional vortex lattice is realized only for the state minimizing the free energy holds for more general vortex lattices or not, we next study 
 another case where the vortex lattice in high fields is well described within the lowest LL. As such a typical case, we consider the following Ginzburg-Landau (GL) model of a superconductor consisting of two-component scalar OPs 
\begin{equation}
{\tilde {\cal H}}_2 = \int d^3\bm{r} \biggl[ - \sum_{j=1,2} \varepsilon_{0,j} |\Delta_j|^2 + \sum_{j=1,2} \xi_0^2 |\bm{\Pi} \Delta_j|^2 + \frac{g}{2} \biggl[ \sum_{j=1,2} |\Delta_j|^4 + \rho \biggl|\sum_{j=1,2} \Delta_j^2 \biggr|^2 \, \biggr] 
+ {\tilde g} |\Delta_1|^2 |\Delta_2|^2 
\biggr] 
\label{GLtriplet}
\end{equation}
expressed by two OP fields $\Delta_s$ ($s=1$, $2$). The vortex lattice structure following from this model in $\rho=0$ case has been studied elsewhere \cite{Chung} as a model appropriate for a nonchiral spin-triplet pairing case. In this section, the vortex lattices and the superfluid stiffness in them following from this model will be examined. 

From the first two terms of (\ref{GLtriplet}), the parameters 
\begin{equation}
\varepsilon_s \equiv \varepsilon_{0,j} - h
\end{equation}
($j=1$, $2$) determining the distance from the $H_{c2}(T)$-curve in the field v.s. temperature phase diagram are defined. As in the preceding section, both of $\varepsilon_1$ and $\varepsilon_2$ are assumed to be positive hereafter. Since we focus on the field range in which the paramagnetic pair-breaking effect is negligible, the coefficient $g$ is always positive, and $\Delta_{s,{\rm MF}}$, which are the mean field solutions of $\Delta_s$, can be assumed to be in the lowest LL at least close to the $H_{c2}(T)$-line. 

Further, to be specific, we focus on the case in which ${\tilde g} \geq 0$, because, in the opposite case with a negative ${\tilde g}$, it is easily understood that the vortices for the two different OPs coalesce to lower the free energy. In contrast, when ${\tilde g} > 0$, the vortices in $\Delta_{1,{\rm MF}}$ and $\Delta_{2,{\rm MF}}$ should be separated from one another to lower the free energy, and hence, the issue \cite{Chung} on what structure of the vortex lattice is realized becomes nontrivial. By representing the separation between two neighboring vortices via $\bm{r}_0$, we set $\Delta_{s,{\rm MF}}$ in the form 
\begin{eqnarray}
\Delta_{1,{\rm MF}} &=& A_0^{(1)} \varphi_0(\bm{r}|0), \nonumber \\
\Delta_{2,{\rm MF}} &=& A_0^{(2)} \varphi_0(\bm{r}|\bm{r}_0), 
\end{eqnarray}
and use the fact that, for the mean field solutions in the lowest LL, the quadratic terms (the sum of the first two terms) of ${\cal H}_2$ may be replaced by $- \sum_{s=1,2} \varepsilon_s |\Delta_{s,{\rm MF}}|^2$. Here, $\varphi_n(\bm{r}|\bm{r}_0)$ is given according to (\ref{nthLL}) and (\ref{Gert}). 

Then, by minimizing ${\cal H}_2$ with respect to $|A_0^{(s)}|^2$, we obtain 
\begin{equation}
\begin{pmatrix}
\varepsilon_1 \\ 
\varepsilon_2
\end{pmatrix}
= g 
\begin{pmatrix} I & J \\
J & I  
\end{pmatrix}
\begin{pmatrix}
|A_0^{(1)}|^2 \\ |A_0^{(2)}|^2
\end{pmatrix},  
\label{MF2}
\end{equation}
where 
\begin{eqnarray}
I &=& (1 + \rho) \langle 0, 0|0, 0 \rangle, \nonumber \\ 
J &=& \frac{{\tilde g}}{g} \, J_0, \nonumber \\
J_0 &=& \langle 0+, 0|0+, 0 \rangle + \frac{\rho}{2} \biggl[ e^{2i\delta_{12}} \langle 0,0|0+, 0+ \rangle + {\rm {c.c.}} \biggr], 
\end{eqnarray}
and the constant $\delta_{12}$, which is the phase of $(A_0^{(1)})^* A_0^{(2)}$, should be determined so that the energy is minimized. 
Here and below, we define 
\begin{eqnarray}
\langle n,m|p,q \rangle \!\!&=&\!\! \langle [ \varphi_n(\bm{r}|0) \varphi_n(\bm{r}|0)]^* \varphi_p(\bm{r}|0) \varphi_q(\bm{r}|0) \rangle_s, \nonumber \\
\langle n+,m|p+,q \rangle \!\!&=&\!\! \langle [ \varphi_n(\bm{r}|\bm{r}_0) \varphi_n(\bm{r}|0)]^* \varphi_p(\bm{r}|\bm{r}_0) \varphi_q(\bm{r}|0) \rangle_s, \nonumber \\
\langle n,m|p+,q+ \rangle \!\!&=&\!\! \langle [ \varphi_n(\bm{r}|0) \varphi_n(\bm{r}|0)]^* \varphi_p(\bm{r}|\bm{r}_0) \varphi_q(\bm{r}|\bm{r}_0) \rangle_s . 
\label{spaceaverages}
\end{eqnarray} 
Then, 
the mean field free energy density $f_{\rm MF}$ becomes 
\begin{equation}
f_{\rm MF} = - \frac{\varepsilon_1^2 + \varepsilon_2^2}{2 g (1 + \rho) 
\beta_A^{(2)}}, 
\end{equation}
where 
\begin{equation}
\beta^{(2)}_A = \frac{1}{1+\rho} \frac{I^2 - J^2}{I - \gamma J}
\label{Abrikosovfactor2}
\end{equation}
with 
\begin{equation}
\gamma = 2 \left(\frac{\varepsilon_1}{\varepsilon_2} + \frac{\varepsilon_2}{\varepsilon_1} \right)^{-1} 
\end{equation}
is the dimensionless parameter determining the lattice structure in the present two-component GL model and corresponding to the Abrikosov factor (\ref{Abrikosovfactor1}) in the single component case. The fact that the variable $\gamma$ depending on the temperature and the field is included in (\ref{Abrikosovfactor2}) implies that the lattice structure in the two-component GL case may change as the temperature or the field is varied. 

\subsection{Structural Phase Diagram}

For later convenience, the parameter 
\begin{equation}
\alpha = \frac{\varepsilon_m}{\varepsilon_M} = \frac{\gamma}{1 + \sqrt{1 - \gamma^2}}   
\end{equation}
will also be defined here, where $\varepsilon_M$ ($\varepsilon_m$) is the larger (smaller) one among $\varepsilon_1$ and $\varepsilon_2$. 
Thus, we only have to find the parameter values of $\bm{r}_0$, $k$, and $\delta_{12}$ minimizing $\beta_A^{(2)}$ under given values of $\alpha$ and ${\tilde g}/{g}$ to determine the lattice structure becoming the mean field solution. 

\begin{table}[htb]
\begin{center}
\small
\begin{tabular}{cccccccccc} \hline
    ${\rm Structure}$ & \vline & ${\rm Rectangle}$ & ${\rm Square}$(1) & ${\rm Square}$(2) & ${\rm Rhombic}$ & ${\rm Triangle}$ & \\ \hline
    $R$ & \vline & 0 & 0 & $0.5$ & $0.5$ & $0.5$ & \\
    $k^2$ & \vline & ${\rm Varied}$ & $2 \pi$ & $\pi$ & ${\rm Varied}$ & $\pi \sqrt{3}$ & 
\\  
    ${\overline y}_0$ & \vline & $k/2$ & $k/2$ & 0 & 0 & $k/3$ 
& \\
\hline
  \end{tabular}
\label{lattice}
\caption{Values of the parameters expressing each structure seen in Figs.2 and 3 are 
summarized. }
\end{center}
\end{table}

%%%%%%%%%%%%%%%%%%% 
\begin{figure}[t]
\scalebox{0.3}[0.3]{\includegraphics{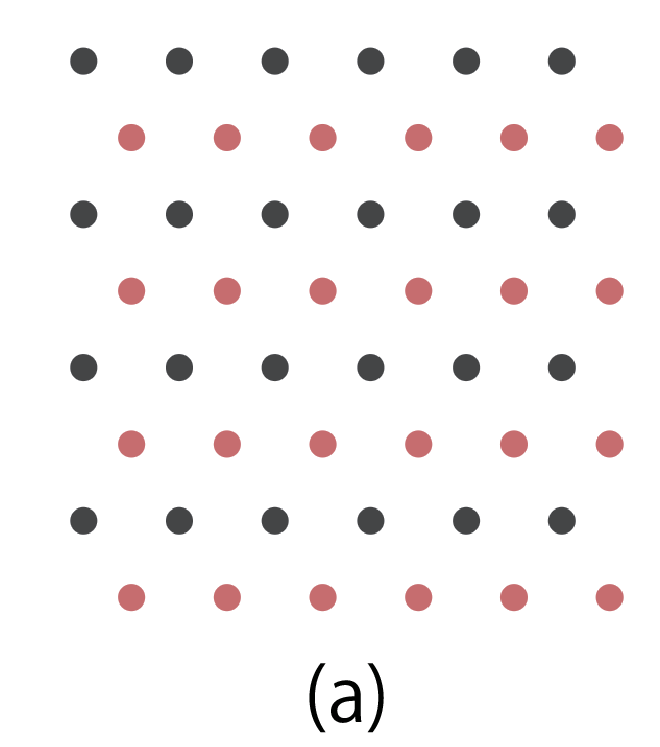}}
\scalebox{0.3}[0.3]{\includegraphics{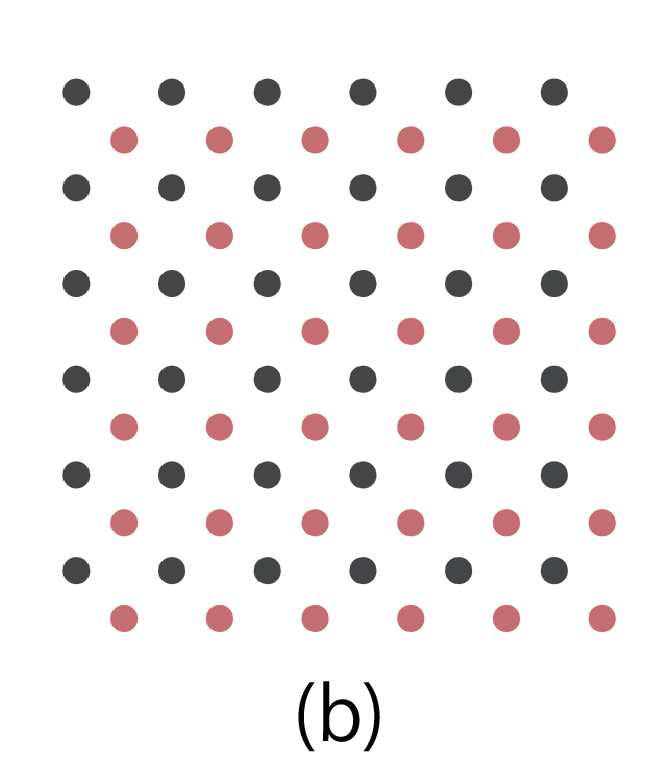}}
\scalebox{0.3}[0.3]{\includegraphics{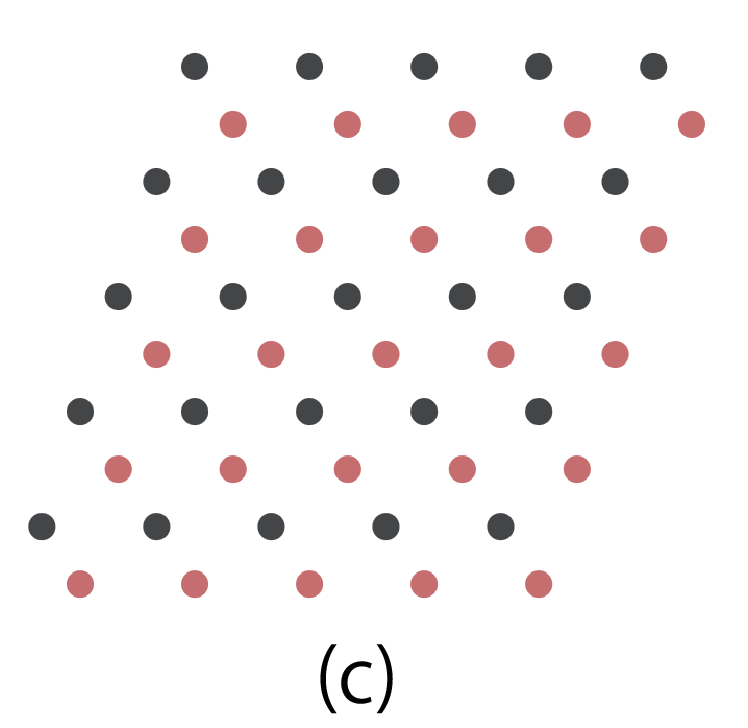}}
\scalebox{0.28}[0.28]{\includegraphics{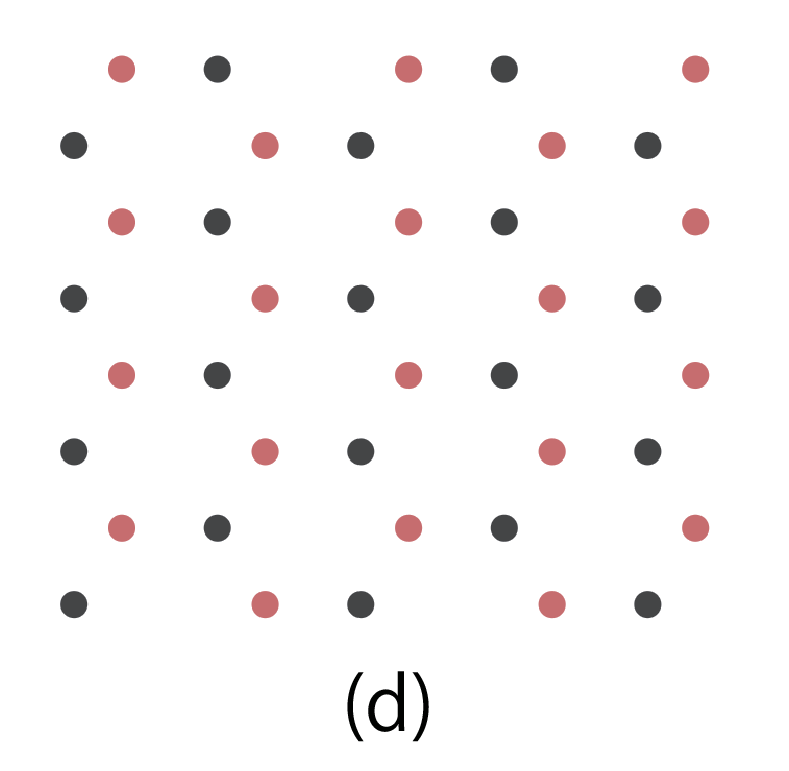}}
\caption{Four types of lattice structures of vortices occurring in the GL model (\ref{GLtriplet}) with $\rho=0$ (see Fig.3) : (a) rectangular (Rec), (b) square (SQ), (c) rhombic (Rh), and triangular (T) lattices. In the figures, the dark green dots are zero points (vortices) of $\Delta_1$, while those of $\Delta_2$ are expressed by light green dots. }
\label{fig.2}
\end{figure}
%%%%%%%%%%%%%%%%%

To find the mean field solution $\Delta_{s,{\rm MF}}$, i.e., the vortex lattice structure minimizing the free energy, for each set of the parameters, $\alpha$ and ${\tilde g}/g$, we have first examined whether a structure with no reflection symmetry is stabilized or not for several values of $\alpha$ and ${\tilde g}/g$ and have found that, under any set of parameter values we have examined, the resulting structure minimizing the free energy has a reflection symmetry. In any case with a reflection symmetry, the parameter $R$ in $\varphi_0(\bm{r}|0)$ can be fixed to be $1/2$ like in the one-component triangular lattice or be zero like in the one-component square lattice. Then, the candidates of the lattice structure minimizing the free energy consist of the structures depicted in Fig.2. The values of the parameters $R$ and $k$ for each structure presented in Fig.2 are shown in Table 1. In all of Fig.2, the dark green dots express the zero points, i.e., the vortex centers, of, say, $\Delta_{1, {\rm MF}}$, while the corresponding zero points of $\Delta_{2, {\rm MF}}$ are indicated by the light green dots. In Fig.2 (a), each of dark green dots and light green ones forms a rectangle lattice, although the entire structure formed by both vortices is a triangular 
one. In the figure (b), the entire structure formed by both of the colored dots is a square lattice as well as each of lattice consisting only of the dark green dots and the one consisting only of the light green ones. The transition between the structures (a) and (b) continuously occurs. The square lattice (b) shows another continuous transition to the rhombic lattice (c). Each rhombus in the structure (c) is continuously varied as the magnetic field is changed. In addition to those three structures, we also have the triangular lattice of the type of the figure (d) in which each light green dot lies at the center of gravity of a triangle formed by three dark green dots and vice versa. 

In this way, in the case with $\rho=0$, we obtain the phase diagram depicted in Fig.3. The obtained sequence of the vortex lattice structures varying with changing the temperature variable $\alpha$ is qualitatively consistent with that reported in Ref.\cite{Chung}. 

Broadly speaking, the structure of the phase diagram in $\rho > 0$ case is qualitatively similar to that in Fig.3. The main differences from those seen in Fig.3 are that the square lattice appeared over a wide parameter range in Fig.3 is replaced by a rhombic one in $\rho > 0$ case, and that the triangular lattice in Fig.3 does not appear any longer for nonzero $\rho$. Except them, the $\alpha$ v.s. ${\tilde g}/g$ phase diagram in $\rho > 0$ case includes, like in Fig.3, several structure transitions. Here, we will not explain further details of the resulting mean field phase diagram in $\rho > 0$ case, because the response of the vortex lattices can be found below irrespective of the $\rho$-value. 

%%%%%%%%%%%%%%%%%%% 
\begin{figure}[t]
\scalebox{0.35}[0.35]{\includegraphics{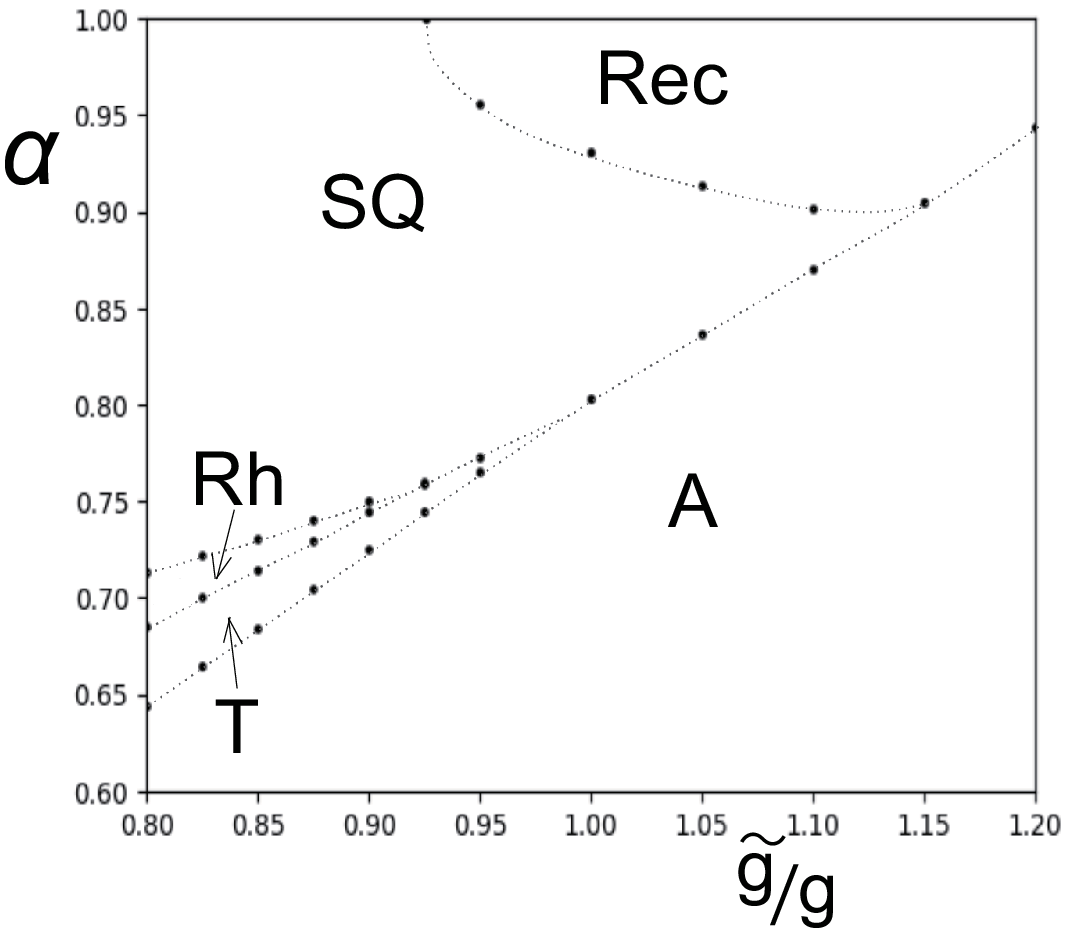}}
\caption{Phase diagram of structural transitions among the vortex lattices expressed in Fig.2 of the model (\ref{GLtriplet}) with $\rho=0$. In the state denoted by A, one of the two OPs vanishes, and the familiar triangular vortex lattice of the nonvanishing OP occurs. }
\label{fig.3}
\end{figure}
%%%%%%%%%%%%%%%%%

\subsection{Superfluid Stiffness} 

Next, we examine the superfluid stiffness $\Upsilon_{ij}$ in the vortex lattices. For this purpose, 
%In this case, one cannot rely on the phase-only approximation because there 
%is no coupling between the two OPs except the ${\tilde g}$ 
%term. 
as well as in the preceding section, we have only to take account of the $n=1$ LL fluctuations for $\Delta_1$ and $\Delta_2$ in the way 
\begin{equation}
\delta \Delta = A_0^{(1)} a_{1,1} \varphi_1(\bm{r}|0) + A_0^{(2)} a_{2,1} \varphi_1(\bm{r}|\bm{r}_0). 
\end{equation}

In the present case, the harmonic fluctuation contribution to ${\cal H}_2$ is given by 
\begin{eqnarray}
\delta {\tilde {\cal H}}_{2} &=& 2h \sum_{s=1,2} |A_0^{(s)}|^2 |{\tilde a}_{s,1}|^2 - \sum_{s=1,2} \varepsilon_s |A_0^{(s)}|^2 |a_{s,1}|^2 + 2 g (1 + \rho) \langle 1,0|1,0 \rangle \sum_{s=1,2} |A_0^{(s)}|^4 |a_{s,1}|^2 \nonumber \\ 
&+& {\tilde g} |A_0^{(1)} A_0^{(2)}|^2 \biggl[ \langle 1+,0|1+,0 \rangle \sum_{s=1,2} |a_{s,1}|^2 + \biggl( \langle 1+,0|0+,1 \rangle + \frac{\rho}{2} [ \langle 1,0|1+, 0+ \rangle e^{2 i \delta_{12}} + {\rm {c.c}} ] \biggr)
(a_{2,1}^* a_{1,1} + {\rm {c.c.}}) \biggr] \nonumber \\ 
&+& 2 {\rm Re} \biggl[ \frac{1+\rho}{2} g \, \langle 0,0|1,1 \rangle \biggl(\sum_{s=1,2} |A_0^{(s)}|^4 a_{s,1}^2 \biggr) + {\tilde g} \, |A_0^{(1)} A_0^{(2)}|^2 \biggl( \langle 0+,0|1+,1 \rangle a_{1,1} a_{2,1} \nonumber \\ 
&+& \frac{\rho}{2} \langle 0+,0+|1,1 \rangle e^{2 i \delta_{12}} [(a_{1,1})^2 + (a_{2,1})^2 ] \biggr) \biggr], 
\label{delH21}
\end{eqnarray}
where 
\begin{equation}
{\tilde a}_{s,1} = a_{s,1} - \frac{1}{\sqrt{2h}} \delta {\tilde A}_+ . 
\end{equation}
We have already used the fact that, just as in the preceding section, the harmonic fluctuation contributions from the gradient term in the original GL term are summarized in the form of the first term of (\ref{delH21}). 
Deleting $\varepsilon_s$ by using eq.(\ref{MF2}), $\delta {\cal H}_2$ is rewritten as 
\begin{eqnarray}
\delta {\tilde {\cal H}}_2 
&=& \sum_{s=1,2} 2h |A_0^{(s)}|^2 |{\tilde a}_{s,1}|^2 
- {\tilde g} |A_0^{(1)} A_0^{(2)}|^2 \biggl( \langle 0+,1|1+,0 \rangle + \frac{\rho}{2} [ \langle 0,0|0+, 0+ \rangle e^{2 i \delta_{12}} + {\rm {c.c}} ] \biggr) \, |{\tilde a}_{1,1} - {\tilde a}_{2,1}|^2 \nonumber \\ 
&+& 2 {\rm Re} \biggl[ \frac{1+\rho}{2} g \langle 0,0|1,1 \rangle (\sum_{s=1,2} |A_0^{(s)}|^4 a_{s,1}^2) + {\tilde g} |A_0^{(1)} A_0^{(2)}|^2 \biggl( \langle 0+,0|1+,1 \rangle a_{1,1} a_{2,1} \nonumber \\ 
&+& \frac{\rho}{2} \langle 0+,0+|1,1 \rangle e^{2 i \delta_{12}} ((a_{1,1})^2 + (a_{2,1})^2) \biggr) \biggr],  
\label{deltaH22}
\end{eqnarray} 
where we have used the relations (\ref{generalbrackets01}) and (\ref{bra1}) 
which are satisfied irrespective of the details of the lattice structure. Further, it will be pointed out in Appendix that, for all structures realized as mean field solutions, the bracket $\langle 1+,0|0+,1 \rangle$ is real and negative. It implies that the equality ${\tilde a}_{1,1} = {\tilde a}_{2,1}$ is favored. 

Based on (\ref{deltaH22}), $\Upsilon_{xy}$ is trivially zero, while $\Upsilon_{xx}$ and $\Upsilon_{yy}$ in high fields, which is valid up to the lowest order in $|\varepsilon_s|/h$, are obtained by setting ${\tilde a}_{s,1}=0$ in eq.(\ref{deltaH22}) and becomes 
\begin{equation}
{\tilde \Upsilon}_{xx} = - {\tilde \Upsilon}_{yy} = \frac{g}{2h} {\rm Re} \biggl[ (1 + \rho) \langle 0,0|1,1 \rangle ( (A_0^{(1)})^4 + (A_0^{(2)})^4 ) + 2 \frac{{\tilde g}}{g}  (A_0^{(1)})^2 (A_0^{(2)})^2 ( \langle +0,0|+1,1 \rangle + \rho \langle 0+,0+|1,1 \rangle e^{2 i \delta_{12}} ) \biggr], 
\label{rostwocomp}
\end{equation}
where ${\tilde \Upsilon}_{ij} = \Upsilon_{ij} \phi_0^2/ [N(0) (2 \pi 
\xi_0)^2]$. 
If using the equality 
\begin{eqnarray}
&-& (I^2 + J^2 - 2 \gamma I J) \partial I +  (2IJ - \gamma(I^2+J^2)) \partial J \nonumber \\ 
&=& (I^2 - J^2)^2 \partial \biggl(\frac{I - \gamma J}{I^2 - J^2} \biggr) , 
\end{eqnarray}
eq.(\ref{MF2}), and the relations in eq.(\ref{derbrackets}) in Appendix, however, one can easily verify 
\begin{equation}
\Upsilon_{xx} = \biggl(\frac{2 \pi \xi_0}{\phi_0} \biggr)^2 \frac{N(0)}{g h (1 + \rho)} (\varepsilon_1^2 + \varepsilon_2^2) \, \partial \biggl(- \frac{1}{\beta_A^{(2)}} \biggr). 
\label{rosresult}
\end{equation}
We note that, if using the conventional notation on the GL model, the prefactor in eq.(\ref{rosresult}) can be connected with the magnetic penetration depth $\lambda(T=0) = \lambda(0)$ in the manner 
\begin{equation}
8 \pi [\lambda(0)]^2 = \frac{g}{N(0)} \, \biggl(\frac{\phi_0}{2 \pi \xi_0} \biggr)^2 . 
\label{londonnotation}
\end{equation}
This result (\ref{rosresult}) is essentially the same as that in the conventional vortex lattice seen in the preceding section in the sense that minimizing the Abrikosov factor, which is $\beta_A^{(2)}$ in the present case, leads to the vanishing of the superfluid stiffness. For brevity, in obtaining (\ref{rostwocomp}), we have neglected higher order corrections in $|\varepsilon_s|/h$. We note that the feature seen above that $\Upsilon_{ii}$ is proportional to $\partial \beta_A^{(2)}$ remains valid even if including such higher order corrections as far as $\Upsilon_{ii}$ is obtained based on (\ref{deltaH22}). 

Since it was found that the superfluid stiffness at zero frequency is zero, the presence of a finite vortex flow conductivity is ensured even for the GL model (\ref{GLtriplet}) consisting of the two-component OPs once dissipative dynamical terms on the OPs are taken into account. Consequences of this result will be mentioned in the final section.

\section{Responses in Vortex Lattice of $D$-wave superconductor} 

In the preceding two sections, we have focused on the superconductors with the field range in which the vortex lattice can be safely described within only the lowest LL subspace of the superconducting OP. When the Cooper-pair wave function has some anisotropy on the Fermi surface of the conduction electrons, however, describing the OP in a field-induced vortex state within only the lowest LL does not become appropriate even just below the $H_{c2}(T)$-line. In fact, in the $d$-wave superconductors, a field-induced rhombic to square transition on the vortex lattice structure originating from the $d$-wave pairing occurs \cite{Ichioka,Huse,Hiasa}, and, for its description, the coupling between the lowest ($n=0$) and $n=4$ LLs needs to be taken into account \cite{Huse,Hiasa}. However, to the best of our knowledge, effects of the $d$-wave pairing-induced anisotropy, i.e., the deviation of the rhombic structure from the six-fold hexagonal symmetry, on the responses and the elasticity of the vortex lattice have not been investigated in the literatures. In this section, by extending the analysis in the preceding sections to the $d$-wave superconductor, the electromagnetic responses and the elasticity of the vortex lattice with the above-mentioned field-induced anisotropy will be examined. 

Since the conductivities under a current perpendicular to the applied magnetic field are also examined below in addition to the static superfluid stiffness, the starting model will be expressed in a form of a quantum action for the appropriate GL Hamiltonian.  The action is expressed by 
\begin{equation}
\frac{\cal S}{N(0) \hbar} = \sum_\omega (\eta|\omega|+ i \eta^{\prime}\omega) \int d^3\bm{r} |\Delta_\omega(\bm{r})|^2 + \frac{1}{\hbar} \int_0^{\hbar \beta} d\tau {\cal H}_d(\Delta(\tau)), 
\end{equation}
where $\eta > 0$, $\Delta(\tau) = (\hbar \beta)^{-1/2} \sum_\omega \Delta_\omega e^{-i\omega \tau}$, $\tau$ is the imaginary time, $\omega$ is the Matsubara frequency for bosons, and 
\begin{eqnarray}
{\cal H}_d(\Delta) &=& \int d^3\bm{r} \biggl[ - \varepsilon_{0} |\Delta|^2 + \xi_0^2 |\bm{\Pi} \Delta|^2 + \gamma \xi_0^4 ( (\Pi_-^2 \Delta)^* (\Pi_+^2 \Delta) \nonumber \\ 
&+& {\rm c.c.} ) + \frac{g}{2} |\Delta|^4 \biggr]. 
\label{GLd}
\end{eqnarray}
Here, $\Pi_+ = \Pi_x - i \Pi_y = \sqrt{2} a^\dagger/r_B + 2 \pi \delta A_-/\phi_0$, and $\Pi_- = \Pi_+^\dagger$. The higher gradient term proportional to $\gamma$ arises from the coupling between the $d_{x^2-y^2}$-pairing function and the spatial gradient. Below, the new dimensionless parameter $\gamma h$ is assumed to be small and will be treated as an expansion parameter as well as $|\varepsilon|/h = (H_{c2}(T) - B)/B$, where $\varepsilon = \varepsilon_0 - h$. 

Using $a$ and $a^\dagger$, defined in Appendix, rather than $\Pi$, (\ref{GLd}) is rewritten as 
\begin{eqnarray}
\frac{{\cal H}_d(\Delta)}{V} &=& - \varepsilon_0 \langle |\Delta|^2 \rangle_s + h \langle \Delta^* (a^\dagger a + a a^\dagger) \Delta  \rangle_s + \langle \delta {\overline A}_+ \delta {\overline A}_- |\Delta|^2 \rangle_s + \sqrt{2h} \langle \delta {\overline A}_+ \Delta^* a^\dagger \Delta + \delta {\overline A}_- \Delta^* a \Delta \rangle_s \nonumber \\ 
&+& 4 \gamma h^2 ( \langle \Delta^* (a^\dagger)^4 \Delta \rangle_s + \langle \Delta^* a^4 \Delta \rangle_s ) + 8 \sqrt{2} \gamma \, h^{3/2} \langle \delta {\overline A}_+ \Delta^* (a^\dagger)^3 \Delta + \delta {\overline A}_- \Delta^* a^3 \Delta \rangle_s + \frac{g}{2} \langle |\Delta|^4 \rangle_s \nonumber \\
&+& 2 \gamma h [ \, \langle (\delta {\overline A}_-)^2 \Delta^* (a^\dagger)^2 \Delta \rangle_s 
+ 4 \langle \delta {\overline A}_- \delta {\overline A}_+ (a \Delta)^* (a^\dagger \Delta) \rangle_s + {\rm c.c.} \, ], 
\label{GLdA}
\end{eqnarray}

As in the preceding sections, the mean field solution of $\Delta$ will be determined within the type II limit and by neglecting its $\tau$-dependence, while the gauge fluctuation $\delta {\overline A}_\pm$ in (\ref{GLdA}) is the external one introduced to see the linear response. It is already known \cite{Lasher} that, in the $\gamma=0$ case, the triangular vortex lattice is formed not only by the $n=0$ LL mode but also by higher $n=6m$ LLs, i.e., with indices of multiples of six. Further, in the present case with the $\gamma$ term, we have the additional expansion parameter $\gamma h$ in addition to $|\varepsilon|/h = (H_{c2} - B)/B$ carried by the $n=6m$ ($> 0$) LLs, and consequently, higher LLs with indices of other even-numbers also participate in the description of the mean field solution $\Delta_{\rm MF}$ even at the lowest order in $(H_{c2} - B)/B$. Concretely, by substituting $\Delta_{\rm MF} = \sum_{m \geq 0} \alpha_{2m} \varphi_{2m}(\bm{r}|0)$ into eq.(\ref{GLdA}) with $\delta {\overline A}_\pm =0$, the mean field free energy density takes the form 
\begin{eqnarray}
\frac{{\cal H}_{d,{\rm MF}}}{V} &=& -\varepsilon |\alpha_0|^2 + \sum_{m \geq 1} (-\varepsilon + 4h m) |\alpha_{2m}|^2 + 4 \gamma h^2 \sum_{m \geq 0} \sqrt{\frac{(2m+4) !}{(2m)!}} (\alpha_{2m}^* \alpha_{2m+4} + {\rm c.c.}) + \frac{g}{2} \biggl[ \langle 0,0|0,0 \rangle |\alpha_0|^4 \nonumber \\ 
&+& 2 |\alpha_0|^2 \sum_{m \geq 1} [\alpha_{2m}^* \alpha_0 \langle 2m,0|0,0 \rangle + {\rm c.c.}] + 2 ( \alpha_2^* \alpha_4^* (\alpha_0)^2 \langle 2,4|0,0 \rangle + {\rm c.c.} ) \nonumber \\ 
&+& 4 |\alpha_0^* \alpha_4|^2 \langle 4,0|4,0 \rangle + \dots \biggr]. 
\label{Fdmf}
\end{eqnarray}

By varying the r.h.s. of eq.(\ref{Fdmf}) with respect to $\alpha_{2m}^*$ and surveying the resulting mean field equations, the dependences on the parameters $\gamma h$ and $|\varepsilon|/h$ of $\alpha_{2m}$ are found as follows : $\alpha_2 = \alpha_0 \,$ O($\gamma h \cdot |\varepsilon|/h$), $\alpha_{4m}$ ($m \neq 3l$) $= \alpha_0 \,$ O($\gamma^m h^m$) or O($\gamma h \cdot |\varepsilon|/h$), $\alpha_{6m} = \alpha_0 \,$ O($|\varepsilon|/h$) (1 + O($\gamma^2 h^2$)), and $\alpha_{10} = \alpha_0 \,$ O($\gamma h \cdot |\varepsilon|/h$). Among them, $\alpha_4$ is the most key quantity induced by the $\gamma$ term of (\ref{GLdA}) reflecting the four-fold symmetry and explicitly given by 
\begin{equation}
\alpha_4 \simeq - \sqrt{6} \alpha_0 \gamma h \biggl( 1 + {\rm O}\biggl(\frac{|\varepsilon|}{h} \biggr) \biggr) . 
\label{alpha4} 
\end{equation}
In obtaining (\ref{alpha4}), we have used the fact that the coupling between the $n=4$ and $n=8$ LLs through the $\gamma$ term leads only to higher order terms in $\gamma h$ and $|\varepsilon|/h$, and consequently that the $n=8$ LL may be neglected from the outset. In contrast, more attention should be paid to the corresponding coupling between the $n=2$ and $n=6$ LLs arising from the $\gamma$ term of (\ref{GLdA}). By using eq.(\ref{alpha4}) and the relations, $\langle 2,4|0,0 \rangle = \sqrt{15} \langle 6,0|0,0 \rangle$ and $\sqrt{2} \langle 0,0|2,0 \rangle = - \langle 0,0|1,1 \rangle$ (see Appendix), one obtains 
\begin{eqnarray}
\alpha_6 &\simeq& - \alpha_0 \frac{g |\alpha_0|^2}{12 h} \langle 6,0|0,0 \rangle \biggl( 1 + {\rm O}( \gamma^2 h^2 ) \biggr), \nonumber \\ 
\alpha_2 &\simeq& \alpha_0 \frac{\sqrt{2} g|\alpha_0|^2}{8h} \langle 1,1|0,0 \rangle - 15 \sqrt{10} \gamma h \alpha_6. 
\label{MFcoff26}
\end{eqnarray}
Although the resulting $\alpha_{10}$ also takes the similar form to 
$\alpha_2$, we do not have to take account of $\alpha_{4m +2}$ with $m \geq 2$. It is because, as is explained later, we focus in the present work on the leading two nontrivial contributions of O($|\alpha_0|^2 \gamma h |\varepsilon|/h$) and O($|\alpha_0|^2 \gamma^2 h^2$) to $\Upsilon_s$, and $\alpha_{10}$ does not contribute to such terms of $\Upsilon_s$. For the same reason, the last line of the expression (\ref{GLdA}), which is proportional to $\gamma h$ and is of O($\delta A^2$), can be neglected from the 
outset. 

Then, as the mean field equation on $\alpha_0$ which is valid up to O($\gamma^2 h^2$), one obtains 
\begin{equation}
\biggl(-\varepsilon + 8 \sqrt{6} \gamma h^2 \frac{\alpha_4}{\alpha_0} \biggr) + g |\alpha_0|^2 {\overline {\langle 0,0|0,0 \rangle}} = 0, 
\label{mfeq}
\end{equation}
where 
\begin{equation}
{\overline {\langle 0,0|0,0 \rangle}} \equiv \langle 0,0|0,0 \rangle + 
24 (\gamma h)^2 \langle 4,0|4,0 \rangle - 4 \sqrt{6} \gamma h \langle 4,0|0,0 \rangle. 
\label{newAbpara}
\end{equation}

%%%%%%%%%%%%%%%%%%% 
\begin{figure}[b]
\scalebox{0.3}[0.3]{\includegraphics{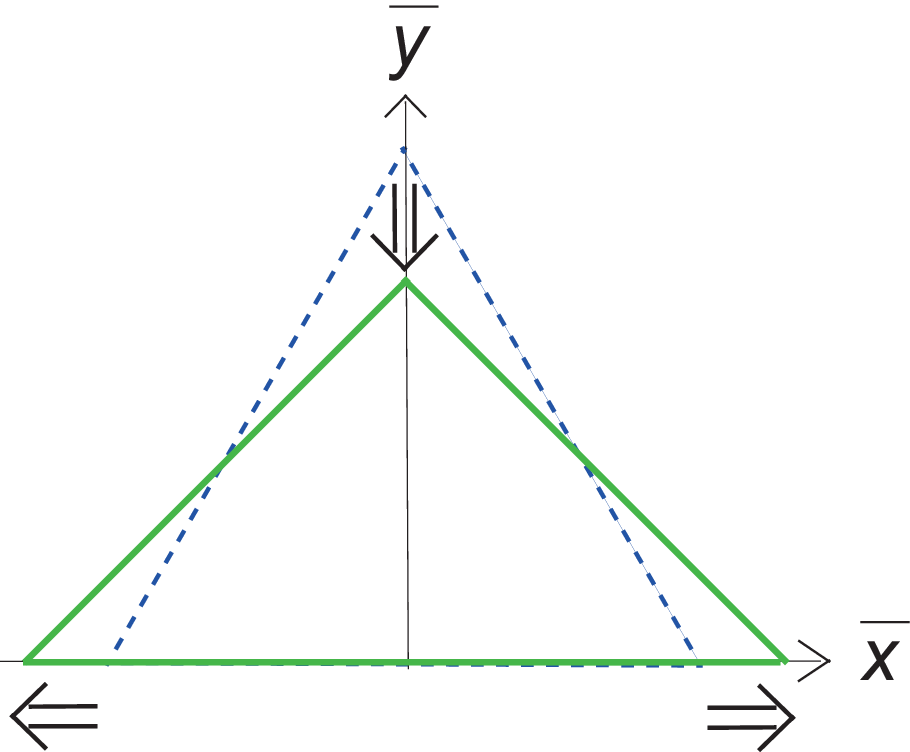}}
\caption{Squashing deformation (black arrows) of a nearly triangular lattice (blue dashed lines) to a square lattice (red solid lines) occurring with increasing the field in the $d$-wave superconductor. }
\label{fig.4}
\end{figure}
%%%%%%%%%%%%%%%%%

By minimizing ${\overline {\langle 0,0|0,0 \rangle}}$ w.r.t. $k^2$, $|\alpha_0|$ is given by 
\begin{equation}
|\alpha_0|^2 = \frac{\varepsilon_0 - h(1 - 48 (\gamma h)^2)}{{\overline {\langle 0,0|0,0 \rangle}}}. 
\label{alpha0}
\end{equation}

Using eq.(\ref{00partial}), minimization of ${\overline {\langle 0,0|0,0 \rangle}}$ implies that 
\begin{equation}
\langle 0,0|1,1 \rangle = 4 \sqrt{6} \gamma h \partial \langle 0,0|4,0 \rangle|_{k=k_0} = 7.48 \gamma h
\label{minimumstructure}
\end{equation}
which is valid up to O($\gamma h$), where $k_0 = \sqrt{\pi \sqrt{3}}$ is the $k$-value of the triangular lattice. Detailed expressions of key brackets appeared above will be presented in Appendix. 

Before moving to our analysis of $\Upsilon_s$ and the conductivity, let us verify that any finite $\gamma$ changes the triangular lattice to a rhombic one. By reexpressing $k^2$ in ${\overline {\langle 0,0|0,0 \rangle}}$ as $k_0^2 + \Delta k^2$, one obtains 
\begin{eqnarray}
{\overline {\langle 0,0|0,0 \rangle}} &=& 1.159595 + \frac{1}{2} \times 0.476 \biggl(\frac{\Delta k^2}{k_0^2} \biggr)^2 \nonumber \\ 
&-& 7.48 \gamma h \frac{\Delta k^2}{k_0^2} + 24 \times \langle 4,0|4,0 \rangle \gamma^2 h^2.  
\label{Abparagamma}
\end{eqnarray} 
The number $0.476$ of the second term of eq.(\ref{Abparagamma}) is familiar as the coefficient accompanying the shear modulus $C_{66}$ of the triangular vortex lattice formed in LLL in the manner \cite{Moore,Labusch} 
\begin{equation}
C_{66} = \frac{g |\alpha_0|^4}{2} 0.476 N(0) . 
\end{equation}
In the present case with a nonzero $\gamma$, (\ref{Abparagamma}) takes the minimum value $-29.33 \gamma^2 h^2$ for $\Delta k^2 = 85.38 \gamma h$. Here, the value in $\gamma \to 0$ limit, $\langle 4,0|4,0 \rangle \simeq 1.222$ (see Appendix), was used. Thus, up to the lowest order in $\gamma h$ and $|\varepsilon|/h$, the vortex lattice is a two-fold symmetric rhombic lattice at any finite field, and the so-called rhombic to square transition is the only field-induced structural transition of the $d$-wave paired vortex 
lattice as far as the crystal anisotropy is neglected \cite{Ichioka,Huse,Hiasa}. 

The action $\delta {\cal S}$ for $\delta {\overline A}_\pm$ and the OP fluctuation $\delta \psi = \sum_{m \geq 0} a_{2m + 1} \varphi_{2m + 1}(\bm{r}|0)$ coupling to $\delta {\overline A}_\pm$ is expressed within the harmonic level in the form 
\begin{eqnarray}
\frac{\delta {\cal S}}{\hbar N(0)} &=& \sum_\omega \biggl[ \, ( |\alpha_0|^2 + |\alpha_4|^2 ) \, \delta {\overline A}_+(\omega) \delta {\overline A}_-(-\omega)  + [ \, \eta|\omega| + i \eta^{\prime} \omega + 2h (1 + 24 \gamma^2 h^2) \, ] |a_1(\omega)|^2 \nonumber \\ 
&+& \!\!\! 8 \sqrt{30} \, \gamma h^2 (a_5^*(\omega) a_1(\omega) + {\rm c.c.}) + \sum_{m \geq 1} \biggl[ ( \eta |\omega| + i \eta^{\prime} \omega + \varepsilon + 2 (2m+1) h) \, |a_{2m+1}(\omega)|^2 \nonumber \\
&+& \!\!\! 4 \gamma h^2 \sqrt{\frac{(2m+5)!}{(2m+1)!}} ( a_{2m+5}^*(\omega) a_{2m+1}(\omega) + {\rm c.c.} ) \biggr] \nonumber \\ 
&+& \sqrt{2h} \biggl[ \delta {\overline A}_+(\omega) \biggl( \sum_{m \geq 0} \biggl[ \sqrt{2m+1} \, \alpha_{2m} a_{2m+1}^*(\omega) + \sqrt{2m+2} \, \alpha_{2m+2}^* a_{2m+1}(-\omega) \biggr] \nonumber \\ 
&+& 8 \gamma h \sum_{m \geq 0} \biggl[ \sqrt{\frac{(2m+3)!}{(2m)!}} \, \alpha_{2m} a_{2m+3}^*(\omega) + \sqrt{\frac{(2m+4)!}{(2m+1)!}} \, \alpha_{2m+4}^* a_{2m+1}(-\omega) \biggr] \biggr) + {\rm c.c.} \biggr] \nonumber \\ 
&+& \frac{g}{2} \biggl[ \biggl( (\alpha_0^*)^2 ( \langle 0,0|1,1 \rangle a_1(\omega) a_1(-\omega) + 2 \langle 0,0|1,5 \rangle a_1(\omega) a_5(-\omega) ) + {\rm c.c.} \biggr) + \dots \biggr] \biggr]. 
\label{GLdfl1}
\end{eqnarray}
In writing (\ref{GLdfl1}), the relation $2 \langle 1,0|1,0 \rangle = \langle 0,0|0,0 \rangle$, eqs.(\ref{mfeq}), and (\ref{newAbpara}) leading to some cancellations between $|a_1|^2$ terms have already been used to rewrite the $|a_1|^2$ terms in the compact form given above. Further, any higher order contributions in $|\varepsilon|/h$ to $\Upsilon_s$ in $\gamma=0$ case has not been taken into account because they have not been included even in the analysis in sec.II. Through the results in sec.II, we already know that, when $\gamma=0$, $\Upsilon_s(\omega=0)$ following from eq.(\ref{GLdfl1}) is zero at least up to the lowest order in $|\varepsilon|/h$. 
Below, we focus on possible lower order terms in $\gamma h$ and $|\varepsilon|/h$ of $\Upsilon_s$, which are of those of O($|\alpha_0|^2 \gamma^2h^2$) of O($|\alpha_0|^2 \gamma h |\varepsilon|/h$). 

In the present case with a finite $\gamma$, there are couplings between the $n$-th LL and $n+4$-th LL. Based on this feature, we first examine effects of the coupling between $n=3$ and $n=7$ LLs. Since the $n=7$ LL fluctuation is accompanied in the terms linear in $\delta \bm{A}$ by $\alpha_6$ and $\alpha_4$, however, it is easily seen that the coupling between $n=3$ and $n=7$ LL fluctuations leads only to higher order contributions to $\Upsilon_s$ than those mentioned in the last paragraph. Then, we only have to freely integrate over $a_3$ to obtain the following contribution to $\delta {\cal S}/(\hbar N(0))$  
\begin{equation}
- \sum_\omega \frac{432 |\alpha_0|^2 \gamma^2 h^2}{6h + \eta|\omega| + i \eta^{\prime} \omega} 
\delta {\overline A}_+(\omega) \delta {\overline A}_-(-\omega). 
\label{a3result}
\end{equation}
It is seen that the $\gamma^2$ dependence of this contribution from the $a_3$-fluctuation arises only from the contribution of the $n=4$ LL to the mean field solution. 

Next, let us move to the contributions of the fluctuations in $n=1$ and $n=5$ LLs. Although, in this case, the $n=9$ LL fluctuation gives an O($\gamma^2h^2$) correction to the "energy gap" $10h$ of the $n=5$ LL fluctuation, this contribution of the $n=9$ LL fluctuation is found not to contribute to the O($|\alpha_0|^2 \gamma^2h^2$) term in the final result of $\Upsilon_s$ at all. Further, since other contributions of the $n=9$ LL fluctuation lead only to higher order corrections in $\gamma h \cdot |\varepsilon|/h$ to $\Upsilon_s$ through the mean field amplitude $\alpha_6$,  $\alpha_8$, and $\alpha_{10}$, we can neglect the $n=9$ LL fluctuation coupling to the $n=5$ LL one from the outset. For convenience of our description, we focus on the particle-hole symmetric case with $\eta^{\prime} = 0$ for the moment. Then, as a result of integrating over $a_5$, the effective form of $\delta {\cal S}$ expressed only by $a_1$ and $\delta \bm{A}$ becomes 
\begin{eqnarray}
\frac{\delta {\cal S}_{\rm eff}}{\hbar N(0)} &=& \sum_\omega \biggl[ \biggl( \eta|\omega|\biggl(1 + \frac{96}{5} \gamma^2 h^2 \biggr) + 2h( 1 - 72 \gamma^2 h^2) \biggr)|a_1(\omega)|^2 - 2h \alpha_0 \alpha_6 \biggl[ \frac{144}{5} \sqrt{5} \gamma h \biggl(1 - \frac{\eta|\omega|}{10 h} \biggr) \nonumber \\ 
&+& 3 \frac{\langle 1,1|0,0 \rangle}{\langle 6,0|0,0 \rangle} \biggr] (a_1(\omega) a_1(-\omega) + {\rm c.c.}) + \sqrt{2h} \biggl[ \delta {\overline A}_+(\omega) \biggl( \alpha_0 \, a_1^*(\omega) \biggl[1 - 72 \gamma^2 h^2 - 12 \gamma^2 h \frac{\eta|\omega|}{5} \biggr] \nonumber \\ 
&+& \alpha_6 \, a_1(-\omega) \biggl[-3 \frac{\langle 1,1|0,0 \rangle}{\langle 6,0|0,0 \rangle} - \sqrt{5} \gamma h \biggl( \frac{138}{5} + \frac{6 \eta |\omega|}{25 h} \biggr) \biggr] \biggr) + {\rm c.c.} \biggr] \nonumber \\ 
&+& |\alpha_0|^2 \biggl( 1 + 3 \gamma^2 h \frac{\eta |\omega|}{5} \biggr) \delta {\overline A}_+(\omega) \delta {\overline A}_-(-\omega) + \frac{6}{5} \sqrt{5} \, \gamma h \, \alpha_0 \alpha_6 \biggl(1 - \frac{\eta|\omega|}{10 h} \biggr) \biggl( \delta {\overline A}_+(\omega) \delta {\overline A}_+(-\omega) + {\rm c.c.} \biggr) \nonumber \\ 
&-& 72 \gamma^2 h^2 |\alpha_0|^2 \biggl(1 - \frac{\eta|\omega|}{6 h} \biggr) \delta {\overline A}_+(\omega) \delta {\overline A}_-(-\omega) 
\biggr]
\label{effaction}
\end{eqnarray}
where the relation $\langle 6,0|0,0 \rangle = - \sqrt{6} \langle 1,5|0,0 \rangle$ following from eq.(\ref{pq00}) was used. The last term is nothing but eq.(\ref{a3result}). Further, the bracket $\langle 1,1|0,0 \rangle$ on the second line followed directly from the $|\Delta|^4$ term, while the corresponding term on the third line occurs from $\alpha_2$ (see eq.(\ref{MFcoff26})). Finally, by integrating over $a_1$, we obtain 
\begin{eqnarray}
F(\delta \bm{A})|_{\eta^{\prime}=0} &\simeq& \frac{g |\alpha_0|^2}{8 \pi [\lambda(0)]^2} \sum_\omega \frac{\eta|\omega|}{2h} \biggl[ (1 + 54 \gamma^2 h^2) (|\delta A_x(\omega)|^2 + |\delta A_y(\omega)|^2) \nonumber \\
&+& 6 \frac{\alpha_6}{\alpha_0} \biggl(12 \sqrt{5} \, \gamma h + \frac{\langle 1,1|0,0 \rangle}{\langle 6,0|0,0 \rangle} \biggr) (|\delta A_x(\omega)|^2 - |\delta A_y(\omega)|^2) \biggr],  
\label{finalrhos}
\end{eqnarray}
where the relation (\ref{londonnotation}) was used. 
This expression (\ref{finalrhos}) vanishing at $\omega=0$ implies that, as expected, the superfluid stiffness is zero. Apparently, this conclusion was obtained without using the condition for the lattice structure minimizing the free energy (\ref{minimumstructure}). In the present case, however, the already-mentioned anisotropy of the vortex lattice structure is measured by the extent of mixing of the third ($n=2$) LL $\varphi_2$ in the mean field solution of the superconducting OP, and hence, optimizing the mixing of the third LL in the mean field solution in part corresponds to the procedure of minimizing the Abrikosov factor $\beta_A^{(1)}$ in sec. II. Therefore, the conclusion in the preceding sections that the vanishing of $\Upsilon_s(\omega=0)$ in an ordered vortex lattice state is a consequence of minimizing the free energy holds even in the present $d$-wave case in which contributions of mixed higher LLs cannot be neglected even close to $H_{c2}(T)$. 

\subsection{Conductivities}

The dc vortex flow conductivity tensor $\sigma_{ij}$ is given by 
\begin{equation}
\sigma_{ij}(\omega = 0) = \frac{\Upsilon_{ij}(\omega)}{\omega} \biggr|_{\omega \to +0}. 
\end{equation}
At a glance, the diagonal (or, dissipative) conductivities $\sigma_{ii}$ ($i=x$, $y$) following from the above expression (\ref{finalrhos}) are expected to be anisotropic so that $\sigma_{xx} \neq \sigma_{yy}$. However, the expression in the parenthesis of the last term of (\ref{finalrhos}) becomes zero, implying that $\sigma_{xx} = \sigma_{yy}$, as a result of minimizing the free energy of the vortex lattice. Therefore, no O($|\alpha_0|^2 \gamma h |\varepsilon|/h$) terms appear in the conductivities, and consequently, the anisotropy of the rhombic lattice structure expressing the deviation from the six-fold symmetric triangular lattice one is not found in the conductivities. That is, we obtain 
\begin{equation}
\sigma_{vf, xx} = \sigma_{bf,yy} = \sigma_{vf}|_{\gamma=0} (1 + 54 \gamma^2 h^2) . 
\end{equation}

So far, we have neglected the $\eta^{\prime}$-term in eq.(\ref{effaction}). As far as focusing on the diagonal and linear conductivities, this procedure is safely valid. Oppositely, in obtaining the linear Hall conductivity arising from a nonzero $\eta^{\prime}$, one may assume $\eta=0$ while keeping a nonvanishing $\eta^{\prime}$. Then, in place of eq.(\ref{finalrhos}), we obtain 
\begin{equation}
F(\delta \bm{A})|_{\eta=0} \simeq \frac{g |\alpha_0|^2}{8 \pi [\lambda(0)]^2} \sum_\omega \frac{\eta^{\prime} \omega}{2h \, B} \, (1 + 6 \gamma^2 h^2) [\delta \bm{A}(-\omega) \times \delta \bm{A}(\omega)]\cdot\bm{B}. 
\end{equation} 
It implies that the vortex flow Hall conductivity is given by 
\begin{equation}
\sigma_{vf,xy} = - \sigma_{vf,yx} = \sigma_{vf,xy}|_{\gamma=0} (1 + 6 \gamma^2 h^2).
\end{equation}
In contrast to the diagonal conductivities, any O($|\alpha_0|^2 \gamma h |\varepsilon|/h$) terms of the Hall conductivity are cancelled with one another irrespective of the minimization of the free energy on the lattice structure. It suggests that the Hall conductivity in the vortex lattice phase is essentially the same as the Hall conductivity obtained for the motion of a single vortex. Further, the correction term proportional to $\gamma^2 h^2$ in $\sigma_{vf,xy}$ has the same sign as the original term (i.e., the first term present even in the $s$-wave case). It implies that no terms induced by the $d$-wave pairing lead to a sign change of the total Hall conductivity. 

\subsection{Tilt moduli}

 As seen in the preceding section, the two-fold anisotropy expressing a deviation of the rhombic lattice from the triangular one is not reflected in the vortex flow conductivity defined in the mean field approximation. However, this anisotropy in the vortex lattice structure arising from the $d$-wave pairing may be reflected at least in a quantity associated with the thermal fluctuation effect. 

To verify this possibility, the tilt moduli appearing in the elastic energy of the rhombic vortex lattice 
will be examined by using the effective action (\ref{effaction}). This can be accomplished by regarding the gauge field $\delta \bm{A}$ there as the internal fluctuation of the flux density. By adding the magnetic energy term and taking account only of the thermal fluctuation contribution with $\omega=0$, we start from the Hamiltonian 
\begin{eqnarray}
{\cal H}_{{\rm eff}} &=& N(0) \int d^3\bm{r} \biggl[ 2h( 1 - 72 \gamma^2 h^2) |a_1(\bm{r})|^2 + 2h \frac{\alpha_6}{\alpha_0} \frac{36}{5} \sqrt{5} \gamma h ( \, (a_1(\bm{r}))^2 + {\rm c.c.}) \nonumber \\ 
&+& \sqrt{2h} \biggl[ \delta {\overline A}_+(\bm{r}) \biggl( \alpha_0 \, a_1^*(\bm{r}) \biggl(1 - 72 \gamma^2 h^2 \biggr) + \alpha_6 \frac{42}{5} \sqrt{5} \gamma h \, a_1(\bm{r}) \biggr) + {\rm c.c.} \biggr] + |\alpha_0|^2 \delta {\overline A}_+(\bm{r}) \delta {\overline A}_-(\bm{r}) \nonumber \\ 
&+& \frac{6}{5} \sqrt{5} \, \gamma h \, \alpha_0 \alpha_6 (\delta {\overline A}_+(\bm{r}) \delta {\overline A}_+(\bm{r}) + {\rm c.c.}) - 72 \gamma^2 h^2 |\alpha_0|^2  \delta {\overline A}_+(\bm{r}) \delta {\overline A}_-(\bm{r})  
\biggr] \nonumber \\ 
&+& \frac{1}{8 \pi} \int d^3\bm{r} (\nabla \times \delta \bm{A})^2, 
\label{effact}
\end{eqnarray}
where the condition (\ref{minimumstructure}) on the equilibrium structure was used. Before proceeding further, the Fourier transformation $a_{1, \bm{q}}$ of $a_1(\bm{r})$ will be identified with the displacement field $\bm{s}^{\rm L}$ of the compressional elastic mode 
\begin{equation}
a_1 = - \alpha_0 \frac{s^{\rm L}_y - i s^{\rm L}_x}{\sqrt{2} r_B}, 
\label{compdisplacement}
\end{equation}
with $(\nabla \times \bm{s}^{\rm L})_z=0$. 

In eq.(\ref{effact}), due to the absence of the $z$-component of the current coupling linearly to the gauge field $\delta \bm{A}$, it is natural to choose the gauge $\delta A_z=0$. By looking at the variational equation with respect to $\delta \bm{A}$, it is found that this choice is equivalent to setting the Coulomb gauge ${\rm div}\delta \bm{A}=0$. Then, by setting $\delta \bm{A} = \nabla \times \varphi(\bm{r}) {\hat z}$ in terms of a scalar field $\varphi$ and integrating over $\varphi$, ${\cal H}_{{\rm eff}}$ is replaced by 
\begin{eqnarray}
E_{\rm comp} &=& \frac{B^2}{8 \pi} \int_{\bm{q}} \frac{1}{D_{\bf q}} \biggl[ q^2 \biggl( 1 - 72 \gamma h \biggl(\gamma h - \frac{\sqrt{5}}{5} \frac{\alpha_6}{\alpha_0} \biggr) \biggr) |s_y(\bm{q})|^2 + q^2 \biggl( 1 - 72 \gamma h \biggl(\gamma h + \frac{\sqrt{5}}{5} \frac{\alpha_6}{\alpha_0} \biggr) \biggr) |s_x(\bm{q})|^2 \nonumber \\
&+& 12 \sqrt{5} \gamma h \frac{\alpha_6}{\alpha_0} \frac{q_x^2 |s_y(\bm{q})|^2 - q_y^2 |s_x(\bm{q})|^2}{\lambda^2 q^2} \biggr], 
\label{elasticenergy1}
\end{eqnarray}
where 
\begin{equation}
D_{\bf q} = \lambda^2 q^2 + 1 - 72 \gamma^2 h^2 + \frac{12}{5} \sqrt{5} \gamma h \frac{\alpha_6}{\alpha_0} \frac{q_y^2 - q_x^2}{q_x^2+q_y^2} , 
\end{equation}
and the relation (\ref{compdisplacement}) was used. As far as the situation dominated by the nonlocal elasticity in which $\lambda^2 q^2 \gg 1$ is concerned, the last term in eq.(\ref{elasticenergy1}) may be neglected. 
Therefore, approximate expressions of the tilt moduli for the displacement in the $i$-th direction ($i=x$, $y$) are 
\begin{eqnarray}
C_{44,x} &=& \frac{B^2}{4 \pi D_{\bf q}} \biggl(1 - 72 \gamma h \biggl(\gamma h + \frac{\sqrt{5} \alpha_6}{5 \alpha_0} \biggr) \biggr), \nonumber \\ 
C_{44,y} &=& \frac{B^2}{4 \pi D_{\bf q}} \biggl(1 - 72 \gamma h \biggl(\gamma h - \frac{\sqrt{5} \alpha_6}{5 \alpha_0} \biggr) \biggr). 
\end{eqnarray}
Further, the interaction range between the vortices in the rhombic vortex lattice is different from the penetration depth $\lambda(T)$ defined in the Meissner state. The interaction ranges $\lambda_i$ ($i=x$, $y$) following from eq.(\ref{elasticenergy1}) are approximately given by 
\begin{eqnarray}
\lambda_x^{-1} &\simeq& \lambda^{-1} \sqrt{ 1 - 72 \gamma^2 h^2 - \frac{12}{5} \sqrt{5} \gamma h \frac{\alpha_6}{\alpha_0} }, \nonumber \\ 
\lambda_y^{-1} &\simeq& \lambda^{-1} \sqrt{ 1 - 72 \gamma^2 h^2 + \frac{12}{5} \sqrt{5} \gamma h \frac{\alpha_6}{\alpha_0} }. 
\end{eqnarray}
Since $\gamma < 0$, and $\alpha_6/\alpha_0 > 0$, the vortices are tilted more easily, and the interaction range is longer in the $y$-direction. 

\section{Summary and Discussion}

In the present paper, we have examined the superfluid stiffness in the mean field vortex lattices occurring in several superconducting systems to clarify the validity of the conventional wisdom that the rigid vortex flow of the vortex lattice is the same as the flow of a single vortex excitation. It has been clarified in the present work that, in the mean field vortex lattice, the vortex flow response occurs only for the lattice structure minimizing the free energy, and hence that the flow motion of a vortex lattice cannot be necessarily identified with a random superposition of single-vortex motions. Further, in the $d$-wave pairing case where the vortex lattice structure in lower fields is a rhombic lattice which is anisotropic compared with the conventional hexagonal one, this anisotropy is found not to be reflected in the conductivity for a current perpendicular to the applied magnetic field ${\bf B}$ as a result of optimizing the vortex lattice structure. 

In sec.III, the conductivities in vortex lattices in the case with multi-component OPs have not been discussed. Since any vortex lattice in this case can be described without any higher LLs like in the $s$-wave conventional case, the results on the conductivities are similar to those in the $s$-wave case. That is, any anisotropy in the resulting vortex lattice structures shown in Fig.2 is not reflected in the conductivities. It does not seem to us that this fact has a common origin to the result in sec.IV that the corresponding anisotropy is not reflected in the conductivities in the $d$-wave case, because the anisotropy of the lattice structure is found not to be reflected in the elastic moduli in the multi-component case, while, as seen in sec.IV, the tilt moduli in $d$-wave case are weakly anisotropic. 

The main conclusion in the present work is that the vanishing of the superfluid stiffness $\Upsilon_{ii}$ ($i=x$, $y$) in an applied magnetic field parallel to $z$-axis is determined by the minimization of the free energy of the vortex lattice structure. There is a possibility that this conclusion may affect theoretical pictures on the pinning effect of the vortex matter, because the elastic theory of the vortex lattice is constructed based on the vanishing of $\Upsilon_{xx}$ and $\Upsilon_{yy}$ \cite{RI95}. Although it is well known that the randomness makes the positional long ranged order of the vortex lattice a short ranged one \cite{Larkin}, it seem that a weaker randomness already makes $\Upsilon_{xx}$ and $\Upsilon_{yy}$ nonvanishing, and consequently, pinning-induced deformations of the vortex lattice are dominated by a plastic one rather than an elastic one. If so, it seems unclear to what extent the so-called collective pinning picture is applicable in real vortex matter. 

\begin{acknowledgement}
The present work was supported by a Grant-in-Aid for Scientific Research [No.21K03468] from the Japan Society for the Promotion of Science. 
\end{acknowledgement}

\vspace{2mm}

* ikeda.ryusuke.5a@kyoto-u.ac.jp

\appendix{\bf {Appendix}}

In this Appendix, several formulae on the brackets to be used in the analysis given in the main text will be derived. The general forms of such brackets are defined in eq.(\ref{spaceaverages}). By imposing the periodic boundary condition and performing the Gauss integral once as in deriving the conventional expression of the Abrikosov factor $\beta_A^{(1)}$, several brackets are written in the following form 
\begin{eqnarray}
\langle 0+,0|0+,0 \rangle &\equiv& \langle (\varphi_0(\bm{r}|\bm{r}_0) \varphi_0(\bm{r}|0))^* \varphi_0(\bm{r}|\bm{r}_0) \varphi_0(\bm{r}|0) \rangle_s \nonumber \\ 
&=& \frac{k}{\sqrt{2 \pi}} \sum_{n,m} \, \exp\biggl(- \frac{k^2}{2}n^2 - \frac{1}{2}(km - y_0)^2 + i (k x_0 - 2 \pi R m)n \biggr), \nonumber \\
\langle 1+,0|0+,1 \rangle &\equiv& \langle (\varphi_1(\bm{r}|\bm{r}_0) \varphi_0(\bm{r}|0))^* \varphi_0(\bm{r}|\bm{r}_0) \varphi_1(\bm{r}|0) \rangle_s \nonumber \\ 
&=& \frac{k}{2 \sqrt{2 \pi}} \sum_{n,m} \, (1 + k^2 n^2 - (km - y_0)^2) \exp\biggl(- \frac{k^2}{2}n^2 - \frac{1}{2}(km - y_0)^2 + i(k x_0 - 2 \pi R m)n \biggr), \nonumber \\ 
\langle 0+,0|1+,1 \rangle &\equiv& \langle (\varphi_0(\bm{r}|\bm{r}_0) \varphi_0(\bm{r}|0))^* \varphi_1(\bm{r}|\bm{r}_0) \varphi_1(\bm{r}|0) \rangle_s \nonumber \\ 
&=& \frac{k}{2 \sqrt{2 \pi}} \sum_{n,m} \, (1 - (km - y_0 - kn)^2) \exp\biggl(- \frac{k^2}{2}n^2 - \frac{1}{2}(km - y_0)^2 + i( k x_0 - 2 \pi R m)n \biggr), \nonumber \\
\langle 1+,0|1+,0 \rangle &=& \langle 0+,0|0+,0 \rangle - \langle 0+,1|1+,0 \rangle. 
\label{generalbrackets01}
\end{eqnarray}
Here, $y_0$ can take any value, while it has been implicitly assumed that $x_0 = 0$ or $\pi/k$. 
From eq.(\ref{generalbrackets01}), the relation 
\begin{equation}
2 \langle 1,0|1,0 \rangle = \langle 0,0|0,0 \rangle, 
\label{bra1}
\end{equation}
straightforwardly follows for any lattice structure. 
Similarly, one can verify that 
the equality 
\begin{equation}
2 \langle 1+,0+|1,0 \rangle = \langle 0+,0+|0,0 \rangle
\end{equation}
is satisfied. 

To make the brackets in eq.(\ref{generalbrackets01}) more useful forms for detailed analysis, those expressions in the two cases with $R=0$ and $R=1/2$ (see Table I) will be written down below individually. 

\subsection{$R=0$} 

The vortex lattices with $R=0$ correspond to the rectangle lattice and the square one appeared in the two-component case in sec.III. Since a vortex for one order parameter lies at the center of the unitcell of the vortex lattice for another order parameter, we will replace $x_0$ and $y_0$ in eq.(\ref{generalbrackets01}) by $\pi/k$ and $k/2$, respectively. Then, it is convenient to rewrite each bracket using the Poisson summation formula 
\begin{equation}
\sum_{n = - \infty}^\infty F(n) = \sum_{m = - \infty}^\infty \int_{-\infty}^\infty dx \, F(x) \, e^{i 2 \pi m x},
\label{poisson}
\end{equation}
where the convergence of the $x$-integral on the function $F(x)$ is assumed. 
For example, $\langle 0,0|0,0 \rangle$ corresponding to the Abrikosov factor $\beta_A^{(1)}$ is rewritten in the way  
\begin{equation}
\langle 0,0|0,0 \rangle = \frac{k}{\sqrt{2 \pi}} \biggl[ \, \sum_m e^{-k^2m^2/2} \, \biggr]^2 
= \sum_n e^{-k^2 n^2/2} \sum_m e^{-2 \pi^2 m^2/k^2}
\end{equation}
as a result of using (\ref{poisson}) once. In the same way, we obtain 
\begin{eqnarray}
\langle 0+,0|0+,0 \rangle &=& \sum_n (-1)^n e^{-k^2 n^2/2} \sum_m (-1)^m e^{-2 \pi^2 m^2/(k^2)}, \nonumber \\ 
\langle 0+,1|1+,0 \rangle &=& \frac{1}{2} \sum_{m,n} (-1)^{m+n} \biggl( \biggl(\frac{2 \pi}{k} m \biggr)^2 + k^2n^2 \biggr) e^{-k^2 n^2/2} e^{- 2 \pi^2 
m^2/k^2}, \nonumber \\ 
\langle 0+,0|1+,1 \rangle &=& \partial \langle 0+,0|0+,0 \rangle, \nonumber \\
\langle 0,0|1,1 \rangle &=& \partial \langle 0,0|0,0 \rangle. 
\label{derbra0}
\end{eqnarray}
The expression of $\langle 1+,0|0+,1 \rangle$ given above suggests that this bracket is negative for $k^2$ of order unity. 
Further, we note that the bracket $\langle 0,0|0+,0+ \rangle$, taking the form 
\begin{equation}
\langle 0,0|0+,0+ \rangle = \frac{k}{\sqrt{2 \pi}} \sum_n (-1)^n e^{-k^2 n_+^2/2} \sum_m (-1)^m e^{-k^2 m_+^2/2}, 
\end{equation}
is clearly zero in this $R=0$ case, where $m_+ (n_+) = m+1/2 (n + 1/2)$. In the same way, it is easily seen that $\langle 0,0|1+,1+ \rangle = \partial \langle 0,0|0+,0+ \rangle$ and $\langle 1,0|1+,0+ \rangle$ are also identically zero in $R=0$ case.

\subsection{$R=\frac{1}{2}$} 

For the brackets for the lattice structures with $R=1/2$ in the two-component superconductor in sec.III, we only have to focus on the use of $x_0=\pi/k$ and $y_0=0$. 
To write down the ensuing expressions of the brackets, 
it is convenient to define 
\begin{eqnarray}
t_{\rm even}^{(n)}(s) &\equiv& \sum_{m={\rm even}} \biggl(\frac{s^2 m^2}{2} \biggr)^n e^{-s^2 m^2/2}, \nonumber \\
t_{\rm odd}^{(n)}(s) &\equiv& \sum_{m={\rm odd}} \biggl(\frac{s^2 m^2}{2} \biggr)^n e^{-s^2 m^2/2}.  
\end{eqnarray}
We note that, in $t_{\rm even}^{(n)}$ ($t_{\rm odd}^{(n)}$), the summation is taken over all even (odd) integers. Then, we have 
\begin{eqnarray}
\langle 0,0|0,0 \rangle &=& t_{\rm even}^{(0)}(k) t_{\rm even}^{(0)}(\pi/k) + t_{\rm odd}^{(0)}(k) t_{\rm odd}^{(0)}(\pi/k), \nonumber \\
\langle 0+,0|0+,0 \rangle &=& t_{\rm even}^{(0)}(k) t_{\rm even}^{(0)}(\pi/k) - t_{\rm odd}^{(0)}(k) t_{\rm odd}^{(0)}(\pi/k), \nonumber \\
\langle 0,0|0+,0+ \rangle &=& t_{\rm even}^{(0)}(k) t_{\rm odd}^{(0)}(\pi/k) - t_{\rm odd}^{(0)}(k) t_{\rm even}^{(0)}(\pi/k), \nonumber \\
\langle 0+,1|1+,0 \rangle &=& t_{\rm even}^{(1)}(k) t_{\rm even}^{(0)}(\pi/k) + t_{\rm even}^{(0)}(k) t_{\rm even}^{(1)}(\pi/k) \nonumber \\ 
&-& t_{\rm odd}^{(1)}(k) t_{\rm odd}^{(0)}(\pi/k) - t_{\rm odd}^{(0)}(k) t_{\rm odd}^{(1)}(\pi/k). 
\label{koushiki1/2}
\end{eqnarray}
All of these expressions follow by using the formula (\ref{poisson}) once. Further, from eq.(\ref{koushiki1/2}), we have 
\begin{eqnarray}
\langle 0,0|1,1 \rangle &=& \partial \langle 0,0|0,0 \rangle, \nonumber \\
\langle 0+,0|1+,1 \rangle &=& \partial \langle 0+,0|0+,0 \rangle, \nonumber \\ 
\langle 0+,0+|1,1 \rangle &=& \partial \langle 0,0|0+,0+ \rangle. 
\label{derbrackets}
\end{eqnarray}
In the case of the square lattice with $k^2 = \pi$, all of the brackets $\langle 0,0|1,1 \rangle$, $\langle 0,0|0+,0+ \rangle$, and $\langle 0+,0|1+,1 \rangle$ vanish. These results are consistent with those seen in $R=0$ case. Note that the square lattice described in terms of $R=0$ and $k^2=2 \pi$ is equivalent to that obtained in terms of $R=1/2$ and $k^2 = \pi$.

Next, as the brackets appearing in the $d$-wave case in sec.IV, the formulae of the brackets including higher LLs will be given which are valid when $R=1/2$. 
In calculating a bracket including higher LLs, it is convenient to use the second representation of eq.(\ref{nthLL}).  
The bracket $\langle p,q|0,0 \rangle$ with positive integers $p$ and $q$ 
becomes
\begin{equation}
\langle p,q|0,0 \rangle = \frac{1}{\sqrt{p!}} \frac{\partial^p}{\partial t_1^p} \frac{1}{\sqrt{q!}} \frac{\partial^q}{\partial t_2^q} \biggl( \sum_{m={\rm even}} \sum_{n={\rm even}} + \sum_{m={\rm odd}} \sum_{n={\rm odd}} \biggr) \exp\biggl( - \frac{k^2 n^2}{2} - \frac{\pi^2 m^2}{2 k^2} - \frac{\sqrt{2}}{2}\biggl(kn + i \frac{\pi}{k}m \biggr)(t_1 - t_2) \biggr) \biggr|_{t_1 \to 0, t_2 \to 0}. 
\label{pq00}
\end{equation}
From this expression, the formulae $\sqrt{2} \langle 2,0|0,0 \rangle = - \langle 1,1|0,0 \rangle$ and $\sqrt{15} \langle 6,0|0,0 \rangle = \langle 2,4|0,0 \rangle$ used in the main text directly follow. 

Applying the Poisson summation formula (\ref{poisson}) to (\ref{pq00}), one obtains 
\begin{eqnarray}
\langle 0,0|2,0 \rangle &=& \frac{1}{\sqrt{2}} \biggl[ \, t_{\rm even}^{(1)}(k) t_{\rm even}^{(0)}(\pi/k) + t_{\rm odd}^{(1)}(k) t_{\rm odd}^{(0)}(\pi/k) - t_{\rm even}^{(0)}(k) t_{\rm even}^{(1)}(\pi/k) - t_{\rm odd}^{(0)}(k) t_{\rm odd}^{(1)}(\pi/k) \, \biggr], \nonumber \\
\langle 0,0|4,0 \rangle &=& \frac{1}{2 \sqrt{6}} \biggl[ \, t_{\rm even}^{(2)}(k) t_{\rm even}^{(0)}(\pi/k) + t_{\rm odd}^{(2)}(k) t_{\rm odd}^{(0)}(\pi/k) + t_{\rm even}^{(0)}(k) t_{\rm even}^{(2)}(\pi/k) + t_{\rm odd}^{(0)}(k) t_{\rm odd}^{(2)}(\pi/k) \nonumber \\ 
&-& 6 \, t_{\rm even}^{(1)}(k) t_{\rm even}^{(1)}(\pi/k) - 6 t_{\rm odd}^{(1)}(k) t_{\rm odd}^{(1)}(\pi/k) \, \biggr], \nonumber \\
\langle 4,0|4,0 \rangle &=& t_{\rm even}^{(0)}(k) t_{\rm even}^{(0)}(\pi/k) + t_{\rm odd}^{(0)}(k) t_{\rm odd}^{(0)}(\pi/k) \nonumber \\ 
&-& 4 \biggl[ \, t_{\rm even}^{(1)}(k) t_{\rm even}^{(0)}(\pi/k) + t_{\rm odd}^{(1)}(k) t_{\rm odd}^{(0)}(\pi/k) + t_{\rm even}^{(0)}(k) t_{\rm even}^{(1)}(\pi/k) + t_{\rm odd}^{(0)}(k) t_{\rm odd}^{(1)}(\pi/k) \, \biggr] \nonumber \\ 
&+& 3 \biggl[ \, t_{\rm even}^{(2)}(k) t_{\rm even}^{(0)}(\pi/k) + t_{\rm odd}^{(2)}(k) t_{\rm odd}^{(0)}(\pi/k) + t_{\rm even}^{(0)}(k) t_{\rm even}^{(2)}(\pi/k) + t_{\rm odd}^{(0)}(k) t_{\rm odd}^{(2)}(\pi/k) \nonumber \\ 
&+& 2 \biggl(t_{\rm even}^{(1)}(k) t_{\rm even}^{(1)}(\pi/k) + t_{\rm odd}^{(1)}(k) t_{\rm odd}^{(1)}(\pi/k) \biggr) \, \biggr] \nonumber \\ 
&-& \frac{2}{3} \biggl[ \, t_{\rm even}^{(3)}(k) t_{\rm even}^{(0)}(\pi/k) + t_{\rm odd}^{(3)}(k) t_{\rm odd}^{(0)}(\pi/k) + t_{\rm even}^{(0)}(k) t_{\rm even}^{(3)}(\pi/k) + t_{\rm odd}^{(0)}(k) t_{\rm odd}^{(3)}(\pi/k) \nonumber \\ 
&+& 3 \biggl(t_{\rm even}^{(2)}(k) t_{\rm even}^{(1)}(\pi/k) + t_{\rm odd}^{(2)}(k) t_{\rm odd}^{(1)}(\pi/k)+ t_{\rm even}^{(1)}(k) t_{\rm even}^{(2)}(\pi/k) + t_{\rm odd}^{(1)}(k) t_{\rm odd}^{(2)}(\pi/k) \biggr) \, \biggr] \nonumber \\ 
&+& \frac{1}{24} \biggl[ \, t_{\rm even}^{(4)}(k) t_{\rm even}^{(0)}(\pi/k) + t_{\rm odd}^{(4)}(k) t_{\rm odd}^{(0)}(\pi/k) + t_{\rm even}^{(0)}(k) t_{\rm even}^{(4)}(\pi/k) + t_{\rm odd}^{(0)}(k) t_{\rm odd}^{(4)}(\pi/k) \nonumber \\ 
&+& 4 \biggl(t_{\rm even}^{(3)}(k) t_{\rm even}^{(1)}(\pi/k) + t_{\rm odd}^{(3)}(k) t_{\rm odd}^{(1)}(\pi/k) + t_{\rm even}^{(1)}(k) t_{\rm even}^{(3)}(\pi/k) + t_{\rm odd}^{(1)}(k) t_{\rm odd}^{(3)}(\pi/k) \biggr) \nonumber \\ 
&+& 6 \biggl(t_{\rm even}^{(2)}(k) t_{\rm even}^{(2)}(\pi/k) + t_{\rm odd}^{(2)}(k) t_{\rm odd}^{(2)}(\pi/k) \biggr) \, \biggr] .
\end{eqnarray}
When $k^2= \pi \sqrt{3}$, one finds that $\langle 0,0|2,0 \rangle = \langle 0,0|4,0 \rangle = 0$ and $\langle 4,0|4,0 \rangle = 1.22$.


\begin{thebibliography}{99}
\bibitem{com} In these references, the vortex flow motion under uniform current had been assumed from the outset to occur.
\end{thebibliography}
\end{document}